\documentclass
[prl,twocolumn,superscriptaddress,floatfix,showpacs,10pt]{revtex4}%
\usepackage{amsfonts}
\usepackage{amsmath}
\usepackage{amssymb}
\usepackage{graphicx}%
\begin{document}
\title{Gravity and the  spin-2 planar Schr\"odinger equation}

\author{Eric A. Bergshoeff}
\email{e.a.bergshoeff@rug.nl}
\affiliation{Centre for Theoretical Physics, University of Groningen, Nijenborgh 4, 9747 AG Groningen, The Netherlands, and 
Institut de Physique Th\'eorique Philippe Meyer,  \'Ecole Normale Sup\'erieure, 
24 Rue Lhomond, 75231 Paris Cedex 05, France.}

\author{Jan Rosseel}
\email{rosseelj@gmail.com }
\affiliation{Faculty of Physics, University of Vienna, Boltzmanngasse 5, A-1090 Vienna, Austria.}

\author{Paul K. Townsend}
\email{p.k.townsend@damtp.cam.ac.uk} 
\affiliation{Department of Applied Mathematics and Theoretical Physics, Centre for Mathematical Sciences, University of Cambridge,
Wilberforce Road, Cambridge, CB3 0WA, U.K.}

\begin{titlepage}
%\begin{flushright}  DAMTP-2017-47 %Preprint-number
%$\hspace{2.1cm}{}$
%\end{flushright}
\vfill

\end{titlepage}

\begin{abstract}

A Schr\"odinger equation proposed for  the GMP gapped spin-2 mode of fractional Quantum Hall  
states is found from a novel  non-relativistic limit, applicable only in  2+1 dimensions, of the massive spin-2 Fierz-Pauli 
field equations.  It is also found from a novel null reduction of the linearized Einstein field equations in 3+1 dimensions, and in this context
a uniform distribution of spin-2 particles implies, via a Brinkmann-wave solution of the non-linear Einstein equations, a confining harmonic
oscillator potential for the individual particles.

\end{abstract}

\pacs{}
\maketitle

\setcounter{equation}{0}

Einstein's theory of General Relativity ceases to be a theory of gravity when considered in a 3D spacetime (i.e. 2+1 dimensions): there is no analog
of the Newtonian force, nor gravitational waves. We call it a ``3D gravity''  theory mainly  because it shares 
with 4D General Relativity  the property of  being  a diffeomorphism invariant theory for a dynamical metric on spacetime, which makes it a useful ``toy model'' for
considering how theories of this type might be compatible with quantum mechanics.

A simple modification of 3D General Relativity  known as ``New Massive Gravity'' (NMG) yields a parity-preserving 3D gravity theory  that {\it does} admit gravitational waves; ``gravitational'' 
in the sense that the corresponding particle excitation of the quantum theory has spin 2, although these spin-2 ``gravitons'' are massive
rather than massless \cite{Bergshoeff:2009hq}. Linearization about a Minkowski vacuum yields a free field theory that is  equivalent to the 3D version of the 
massive spin-2 field theory  proposed long ago by Fierz and Pauli \cite{Fierz:1939ix}.  There are various {\it bi-metric} 3D gravity theories that have 
the same linearized  limit \cite{Banados:2009it,Bergshoeff:2013xma},  and NMG may itself  be viewed as the simplest example, with an auxiliary tensor field 
as the second `metric' \cite{Bergshoeff:2009aq}. 

Although NMG has no ``real-world''  applications {\it as a theory of gravity}, it has potential applications in the world of condensed matter systems in 2+1 dimensions. 
Naturally, these are typically non-relativistic, so this motivates consideration of the non-relativistic limit of NMG. Non-relativistic limits are notoriously more 
complicated than one would naively imagine, so it makes sense to first investigate the non-relativistic limit of the 3D  Fierz-Pauli (FP) theory.  One might
expect to find  a Schr\"odinger equation for a non-relativistic particle of spin 2. 

As it happens, fractional Quantum Hall states have  a Girvin-MacDonald-Platzman (GMP) gapped spin-2 mode \cite{GMP}, and a particular Schr\"odinger equation 
has been proposed as an equation governing its dynamics \cite{MHKPS,YCF}.  Following suggestions of  a geometrical interpretation of GMP states \cite{Haldane:2011ia}, this 
spin-2 planar Schr\"odinger equation was shown to emerge upon linearization of a particular  non-relativistic bi-metric theory \cite{Gromov:2017gsk,Gromov:2017qeb}. 
We should stress that these are {\it space} metrics rather than spacetime metrics, but an obvious question is whether this
bi-metric theory is the non-relativistic limit of some relativistic bi-metric theory, perhaps NMG.  We do not answer this question, but we  show that 
the Schr\"odinger equation proposed to describe the GMP mode is indeed a non-relativistic limit of the 3D FP theory. 

The standard way in which spin is incorporated into the (time-dependent)  Schr\"odinger equation is via a multiplet of  complex wavefunctions transforming in a representation
 of the rotation group.  This  implies an $SO(2)$ doublet for two space dimensions,  and  this  is indeed what one finds from the standard non-relativistic limit of  the 3D FP 
 theory for a {\it complex} tensor field, but  the ``spin-2 Schr\"odinger equation'' proposed to describe the GMP mode has a {\it single} complex wavefunction.  What we need,  although only for 2+1 dimensions, is a non-relativistic limit for a {\it real} FP tensor field.

There is a problem with the non-relativistic limit of real-field theories that propagate massive modes. This can be understood
by considering the Klein-Gordon (KG) equation for a scalar field $\Phi$ of mass $m$. Including all factors of $c$ and $\hbar$ this is
\begin{equation}
\frac{1}{c^2} \ddot\Phi - \nabla^2\Phi + \left(\frac{mc}{\hbar}\right)^2 \Phi=0\, . 
\end{equation}
The $c\to\infty$ limit can be taken directly provided  that the reduced Compton wavelength $\lambda= \hbar/(mc)$ is held fixed, but this yields 
a Yukawa equation (Laplace if $1/\lambda=0$), which is non-dynamical. However,  {\it if $\Phi$ is complex} we can set
\begin{equation}
\Phi = e^{- \frac{i}{\hbar}(mc^2-E_0)  t} \Psi \, ,
\end{equation}
where $E_0$ is constant and $\Psi$ is a new complex scalar field. The KG equation becomes
\begin{equation}\label{KG0}
-\frac{1}{2mc^2} \Big(i\hbar \frac{d}{dt} -E_0\Big)^2\Psi  -i\hbar\dot\Psi  - \frac{\hbar^2}{2m} \nabla^2\Psi + E_0 \Psi=0\, , 
\end{equation}
and the $c\to\infty$ limit yields the Schr\"odinger equation
\begin{equation}
i\hbar\dot\Psi = H \Psi\, , \qquad H=  - \frac{\hbar^2}{2m} \nabla^2 + E_0\, . 
\end{equation}
Clearly, this procedure is not applicable for a real scalar field, and there is a group theoretical reason for this difficulty.  The Bargmann symmetry group  
of the Schr\"odinger equation  has one more generator than the Lorentz symmetry group of the KG equation, a central charge proportional  to the mass $m$. This implies that the wavefunction provides only a projective representation of the Galilei group, so it must be complex, and hence the initial KG field must also be complex. The KG equation then has an additional $U(1)$ phase invariance, so there is no longer a mismatch in the dimension of the relativistic and non-relativistic symmetry groups. 
\bigskip

\noindent
{\bf A new non-relativistic limit.} The same difficulty applies to real tensor fields, such as the symmetric traceless tensor field $f_{\mu\nu}$ of the spin-2 FP equations; traceless
 in the sense that  $\eta^{\mu\nu}f_{\mu\nu} =0$, where $\eta^{\mu\nu}$ is the inverse of the background Minkowski metric tensor.   The FP equations comprise  second-order ``dynamical'' 
 equations and first-order ``subsidiary conditions'':
\begin{equation}\label{FPeqs}
\left[\square -(mc)^2\right] f_{\mu\nu} =0 \, , \qquad \eta^{\mu\nu}\partial_\mu f_{\nu\rho} =0\, , 
\end{equation}
where $\square \equiv \eta^{\mu\nu}\partial_\mu\partial_\nu$.  Here  we set  $\hbar=1$, in which case $mc$ has dimensions of inverse length. 
Although the standard path to a non-relativistic limit of these equations requires $f_{\mu\nu}$ to be complex, another non-relativistic limit is possible for 
a Minkowski background of 2+1 dimensions.  In this 3D case we have $\mu,\nu=0,1,2$,  the  Minkowski metric matrix is ${\rm diag.}(-c^2,1,1)$, and 
$f_{\mu\nu}$ has five independent components parametrizing a scalar, vector  and traceless symmetric tensor  of the $SO(2)$ rotation group. The scalar is the real variable
\begin{equation}
f_{00}\equiv c^2(f_{11}+f_{22}) \, , 
\end{equation}
while the vector and traceless symmetric tensor are, respectively, the complex variables
\begin{equation}
f[1] = f_{01} + if_{02}\, , \qquad f[2] = \frac{1}{2}\left(f_{11}-f_{22}\right) + if_{12}\, . 
\end{equation}
In terms of these variables, the subsidiary conditions are
\begin{equation}
\dot f_{00}  = c^2 \Re \left[\bar\partial f[1]\right]\, , \quad \dot f[1] = c^2\bar\partial f[2] + \frac{1}{2}\partial f_{00}\, . 
\end{equation}
As both $f[1]$ and $f[2]$ are complex, we may  set
\begin{equation}\label{NRstep}
f[n] = e^{-i(mc^2-E_0) t} \Psi[n]\, ,  \qquad n=1,2, 
\end{equation}
for new complex variables $\Psi[n]$. The dynamical equations for $f[n]$ are then solved to leading order as $c\to\infty$ if $\Psi[n]$
and $\dot\Psi[n]$ remain finite in this limit.  Given this, the subsidiary  conditions imply that 
\begin{eqnarray}\label{subsid2}
f_{00} &=& - m^{-1} \Im (\bar\partial f[1]) + {\cal O}\left(1/c^2\right) \, , \nonumber \\
f[1] &=& im^{-1} \bar\partial f[2] + {\cal O}\left(1/c^2\right) \, . 
\end{eqnarray}
Only $f[2]$ is independent, and its dynamical equation is 
\begin{equation}\label{dynamic2}
\frac{1}{c^2} \ddot f[2] + \left[(mc)^2 - \nabla^2\right] f[2] =0\, . 
\end{equation}
In terms of $\Psi[2]$ this equation takes the form (\ref{KG0}) and its $c\to\infty$ limit  is
\begin{equation}\label{Schr2}
i\dot\Psi[2] = H\Psi[2] \, , \qquad H= -\frac{1}{2m} \nabla^2 + E_0\, . 
\end{equation}
This Schr\"odinger equation is  invariant under a symmetry group with one more generator than the Lorentz invariance group 
from which we started because the orbital angular momentum  and the spin angular momentum  are {\it separately} conserved in the $c\to\infty$  limit, 
with a ``spin rotation'' becoming  a phase rotation by double the angle.  In fact, this Schr\"odinger equation  is identical 
to the Schr\"odinger equation for  the spin-2 GMP mode of fractional Quantum Hall states, as deduced in \cite{Gromov:2017qeb} from a 
{\it non-relativistic} bi-metric theory. 
\bigskip

\noindent
{\bf Parity and Time-reversal.}  As mentioned earlier, the FP equations for a {\it complex} tensor field allow a standard non-relativistic limit. This leads to
a parity-preserving  {\it pair} of Schr\"odinger equations:
\begin{equation}
i\dot \Psi[\pm 2] = H \Psi[\pm2]\, , \qquad \Psi[\pm 2] = \varphi_{11} \pm i \varphi_{12}\, ,
\end{equation}
where $\varphi_{11}$ and $\varphi_{12}$ are the independent {\it complex} components of a complex symmetric traceless 2-space tensor $\varphi_{ij}$.  The  wavefunctions $\Psi[\pm2]$ 
are spin-2 helicity eigenstates; by ``helicity'' we mean the spin angular momentum while  ``spin'' is  its absolute value.  The point of this discussion is that parity reverses the sign of helicity, and hence exchanges $\Psi[2]$ with $\Psi[-2]$. This can be seen from the following equivalence 
\begin{equation}\label{selfd}
\varphi_{ij} =  \pm i \epsilon_{ik} \varphi_{kj}\ \Leftrightarrow \ \Psi[\mp2] =0\, . 
\end{equation}
If we choose to impose this constraint with the upper sign then we are left with the single Schr\"odinger equation of (\ref{Schr2}). Furthermore, 
the condition $\Psi[-2]=0$ identifies a a spin-rotation of angle $\theta$ with a shift of the phase of $\Psi[2]$ by angle $2\theta$, exactly as required. 

As the real-field FP equations (\ref{FPeqs}) are parity invariant, it follows that  our ``new non-relativistic limit''  of these equations must break parity.  
To see why,  we observe that parity for the  FP equations takes $f[2]\to \bar f[2]$, but one can see from  (\ref{NRstep}) that the corresponding transformation of 
$\Psi[2]$ is not defined in the $c\to\infty$ limit.   In contrast, time-reversal, which takes $t\to-t$ {\it and} $f[2]\to \bar f[2]$ 
(since its action is anti-linear)  {\it is} defined in the  $c\to\infty$ limit, so  (\ref{Schr2}) must be, and is,  time-reversal invariant.

The fact that (\ref{Schr2}) breaks parity suggests that a better starting point for this paper might have been the  parity violating equations of 
``Topologically Massive Gravity''  (TMG) \cite{Deser:1981wh}, which has the ``square-root FP'' equations \cite{Aragone:1986hm} as its linearized limit.   
Moreover, the self-duality condition (\ref{selfd}) (for one choice of the sign) emerges naturally from the non-relativistic limit of the {\it complexified} ``square-root FP'' 
equations, as detailed for the spin-1 case  in {\cite{Bergshoeff:2018tjg}. However, it is not clear how  the required ``complexification'' is to be implemented for the
non-linear TMG theory. This is not a problem for NMG because complexification is not required. 
\bigskip

\noindent
{\bf Generalized null reduction.} We now turn to a different derivation of the Schr\"odinger  equation (\ref{Schr2}).  It is well-known that null reduction of a Lorentz invariant theory in a 
5D Minkowski spacetime yields a Galilean invariant theory in 1+3 dimensions \cite{Gomis:1978mv,Duval:1984cj,Julia:1994bs}. We seek some variant procedure that will 
take the linearized 4D Einstein field equations  to the Schr\"odinger equation (\ref{Schr2}). A similar issue was addressed in \cite{Cariglia:2016oft} at the level of particle mechanics:
the Hamilton-Jacobi equation for a non-relativistic particle in $d$ space dimensions was provided with an ``Eisenhart lift'' to $d+1$ dimensions.  Here we propose a quantum 
version in which the planar Schr\"odinger equation (\ref{Schr2}) is lifted to the linearized 4D Einstein equations; in reverse this becomes a generalized null reduction inspired by 
Scherk-Schwarz dimensional reduction \cite{Scherk:1978ta}.  In principle the idea applies in any dimension but it is only for a {\it planar} Schr\"odinger equation that one can lift to
the {\it real-field} linearized Einstein equations. 

Linearization of the 4D vacuum Einstein equations about a  Minkowski vacuum with coordinates $\{x^m; m=0,1,2,3\}$ and Minkowski metric $\eta_{mn}$, yields the following equations for the metric perturbation tensor:
\begin{equation}\label{linE}
\square h_{mn} -2\partial_{(m}h_{n)} + \partial_m\partial_n h =0\, , 
\end{equation}
where $h_m \equiv \eta^{pq}\partial_p h_{qm}$ and $h\equiv \eta^{mn}h_{mn}$. 
We shall choose light-cone coordinates for which $x^m= \{x^+,x^-, x^i\}$, where $i=1,2$ and $x^\pm = (x^3\pm x^0)/\sqrt{2}$, 
and units for which $c=1$, but we no longer set $\hbar=1$.

The standard null reduction is achieved by requiring $\partial_-h_{mn}=0$.  Instead, we proceed on the assumption that  $\partial_-$ is invertible, in which case 
we may impose the light-cone gauge condition $h_{m-}=0$, for which 
\begin{eqnarray}\label{hs}
h &=& h_{ii} \, , \quad h_+ = \partial_- h_{++} + \partial_i h_{i+} \, , \nonumber \\
h_- &=&0\, , \qquad h_i  = \partial_- h_{i+} + \partial_j h_{ij} \, . 
\end{eqnarray}
The equation for $h_{m-}$ reduces to $\partial_- (h_m - \partial_m h)=0$, which implies that $h_m=0$ and $h=0$; these equations imply that 
\begin{equation}\label{auxils}
\partial_- h_{++} = -\partial_ih_{i+} \, , \qquad \partial_- h_{i+} = - \partial_j h_{ij}\, , 
\end{equation}
and also that the linearized Einstein equations reduce to $\square h_{mn}=0$.  As $\partial_-$ is assumed invertible, we may solve for 
the auxiliary variables $h_{i+}$ and $h_{++}$. This leaves only the traceless part of  $h_{ij}$, which satisfies the 4D wave equation, 
implying the propagation of transverse waves with two independent polarizations. So far, this is standard light-cone gauge fixing. 

Next,  we define 
\begin{equation}\label{defs}
\Psi[1] = h_{1+}+ ih_{2+}\, , \qquad  \Psi[2]= h_{11} + ih_{12}\, . 
\end{equation}
The auxiliary variable equations (\ref{auxils}) are now
\begin{equation}\label{auxiliary}
\partial_- h_{++} = -\Re\left( \bar\partial \Psi[1]\right) \, , \qquad  \partial_- \Psi[1] = - \bar\partial \Psi[2]\, ,  
\end{equation}
and the wave equation for the traceless transverse metric perturbation is 
\begin{equation}
2\partial_- \partial_+ \Psi[2] = - \nabla^2 \Psi[2]\, . 
\end{equation}

We now propose to effect a new null reduction  by setting 
\begin{equation}\label{genred}
\partial_- \Psi[n]= i(m/\hbar) \Psi[n]\, ,  \quad n=1,2\, ,  
\end{equation}
for positive mass $m$; the factor of $\hbar$ is needed here on dimensional grounds.  The equations (\ref{auxiliary}) now imply that 
\begin{eqnarray}\label{solved}
h_{++} &=& - (\hbar/m) \Im \left(\bar\partial\Psi[1]\right) + {\rm const.} \nonumber \\
\Psi[1] &=&  i(\hbar/m) \bar\partial\Psi[2] \, . 
\end{eqnarray}
These equations are analogous to the subsidiary equations for the 3D spin-2 FP equations in the form 
(\ref{subsid2}).  As in that case, only  $\Psi[2]$ is independent, and it satisfies 
\begin{equation}\label{Schr3}
  -\frac{\hbar^2}{2m} \nabla^2 \Psi[2] = i\hbar\dot\Psi[2]\, , 
\end{equation}
where $\dot\Psi \equiv \partial_+\Psi$. This is the Schr\"odinger equation  (\ref{Schr2}), with $E_0=0$ but we address this below.

One may again ask how it is that the parity invariance of our starting point is not reflected in the end result, and in this case the answer is that
parity is broken  by the choice of sign for the mass $m$ appearing in (\ref{genred}). If we had supposed  $m$ to be negative then we would have had to take 
the complex conjugate of (\ref{Schr3}) to arrive at a standard Sch\"odinger equation for $\bar\Psi[2]$, but it follows from the definition of $\Psi[2]$ in (\ref{defs}) 
that $\bar\Psi[2]= \Psi[-2]$.

We have now provided two distinct ``gravity'' interpretations of the planar Schr\"odinger equation that has appeared in the context of the spin-2 GMP mode of fractional Quantum Hall states.   Our  interpretation  of  it  as a non-relativistic limit 
of the 3D FP equations is closest to   ``geometrical''  proposals  in the condensed-matter literature, but  our derivation from the 4D linearized 
Einstein equations provides a more direct link to ``gravity''. In both cases the enabling  feature is the fact 
that the relevant subgroup of the Lorentz group is $U(1)$; this is the rotation group in 3D and the transverse rotation group in 4D, and 
Wigner's ``little group''  in both cases because in 3D the spin-2 particle is massive whereas in 4D it is massless. 
\bigskip

{\bf Gravity and the Schr\"odinger potential.} So far, we have discussed the planar Schr\"odinger equation only for a {\it free} spin-2 particle, 
or spin-2 GMP mode in the condensed matter context. The gravity origin of this equation becomes useful when we consider how these particles
might interact. In this relativistic context, each particle will produce a gravitational field that is felt by all the others, and we can approximate the 
effect on any individual particle by some collective background spacetime metric; each particle then moves freely in this background. We may 
anticipate that this mean-field type of approximation will result in some potential for the Schr\"odinger Hamiltonian.

To explore this idea in the context of generalized null-reduction, we must start from some solution of the full 4D Einstein field equations: 
$G_{mn} = 8\pi G_NT_{mn}$,  where $G_{mn}$ is the Einstein tensor, $G_N$ is Newton's constant, and $T_{mn}$ is some specified source 
tensor (we again set $c=1$).  Given a 4-metric that solves these equations, we  linearize about it to find the following equations 
for the metric perturbation tensor: 
\begin{eqnarray}\label{lineq2}
0&=& D^2 h_{mn} - 2D_{(m} h_{n)} + D_m \partial_n h \nonumber \\
&&+ \ 2\left[R_{p(m}h_{n)}{}^p + R_{pmnq} h^{pq}\right]\,  , 
\end{eqnarray}
where $D$ is the covariant derivative with respect to the affine connection for which $R^p{}_{mnq}$ is the Riemann tensor and
$R_{mn}$ the Ricci tensor,  and 
\begin{equation}
h_m = g^{pq}D_ph_{qm}\, , \quad h= g^{mn}h_{mn}\, . 
\end{equation}

We choose as our background the particular Brinkmann-wave metric
\begin{equation}
ds^2 = 2dx^+dx^- + 2v(x^+,{\bf x}) (dx^+)^2 + d{\bf x}\cdot d{\bf x}\, , 
\end{equation}
which also played a role in the ``Eisenhart lift'' of  \cite{Cariglia:2016oft}, and earlier in \cite{Gibbons:2010fb}.

The function $v$ is independent of $x^-$, which  ensures that $\partial_-$ is a null Killing vector field; in particular, 
constant $v$ yields Minkowski spacetime.   The only non-zero components of the affine connection, up to symmetry, are
\begin{equation}
\Gamma_{++}{}^- = \partial_+ v\, , \quad \Gamma_{++}{}^i =  -\partial_i v\, , \quad \Gamma_{i+}{}^- = \partial_i v\, , 
\end{equation}
and the only non-zero components of the curvature and Ricci tensors, up to symmetries, are
\begin{equation}
R_{+ij+} = \partial_i\partial_j v\, , \qquad R_{++} =  \nabla^2 v\, ,  
\end{equation}
where $\nabla^2$ is the Laplacian on the transverse 2-space. 
The background Einstein equations are satisfied if 
\begin{equation}\label{Poisson}
\nabla^2 v = 8\pi G_N T_{++}\, , 
\end{equation}
where $T_{++}$ must be the {\it only} non-zero component of $T_{mn}$.  

As before, we impose the gauge condition $h_{m-}=0$. The Ricci tensor term in (\ref{lineq2}) is then zero. The function $v$ does not enter into 
the expressions for $h_m$ and $h$ in light-cone gauge, and neither does it enter into the  ``dynamical'' equation for $h_{m-}$, so we still have  $h_m=h=0$, 
and the resulting equations (\ref{auxils}), while the dynamical equations reduce to 
\begin{equation}
D^2h_{mn} + 2R_{mpqn}h^{pq}  =0\, . 
\end{equation}
We need consider only the equation for $h_{ij}$, which is 
\begin{equation}
2\partial_-\partial_+h_{ij} - 2v \partial_-^2 h_{ij} + \nabla^2 h_{ij} =0\, . 
\end{equation}
Only the traceless part of $h_{ij}$ is non-zero, and we can trade this for $\Psi[2]$ as before. 
Imposing the generalized null-reduction condition (\ref{genred}),  we again recover the equations (\ref{solved}) determining the auxiliary fields in terms of $\Psi[2]$, while the equation for $\Psi[2]$
again becomes the spin-2 planar Schr\"odinger equation,  but now with Hamiltonian 
\begin{equation}
H= -\frac{\hbar^2}{2m} \nabla^2 + V(t,{\bf x})\, , \qquad V=mv\, , 
\end{equation}
where $t=x^+$. 
One  solution of (\ref{Poisson}) for  zero source yields the linear potential $V= m{\bf g}\cdot{\bf x}$, which is naturally interpreted as the result of a constant acceleration ${\bf g}$.   

The simplest non-zero source is $T_{++} = \rho$, for  {\it constant} $\rho$.  In this case the general rotationally invariant solution of (\ref{Poisson}) for positive
$v$ is  \begin{equation}
v= \frac{1}{2} \omega^2 |{\bf x}- {\bf x}_0|^2\, ,  \qquad \omega^2 = 8\pi G_N\rho\, . 
\end{equation}
This yields the Hamiltonian for a  planar harmonic oscillator of angular frequency $\omega$, which confines the particle to a region  centered on the  arbitrary point with coordinates ${\bf x}_0$.  The natural interpretation is that of a constant uniform distribution of spin-2 particles with each particle  occupying an area $\hbar/(m\omega)$. 

An obvious  question is whether this result can also be found from the non-relativistic limit of some interacting extension of the 3D spin-2 FP theory, such as NMG. 
In this context, the potential has an interpretation within Newton-Cartan geometry as the time component of the gauge-potential $1$-form associated to the 
central-charge of the Bargmann algebra \cite{Andringa:2010it,Andringa:2013mma}; a gauge transformation preserving the form of this potential shifts $v$ by a function of $t$, which  corresponds to the freedom 
to redefine the wavefunction by a $t$-dependent phase factor.  However, it is not clear to us at present how this modification can be implemented  in the context of the 
new non-relativistic limit described here that avoids complexification of the FP field. 

Finally, we should mention that   a study by Vasiliev  \cite{Vasiliev:2012vf} of relativistic conformal field theories in their ``unfolded'' formulation led to a
holographic-dual  Schr\"odinger equation in one lower dimension, and a proposed twistor transform interpretation that was conjectured to be related
to what we have called, following \cite{Cariglia:2016oft,Gibbons:2010fb}, the ``Eisenhart lift''.

\bigskip

\noindent\textbf{Acknowledgements}:  We thank Gary Gibbons, Andrey Gromov and  Dam T. Son for discussions,  and the ENS for hospitality. 
The work of PKT is partially supported by the STFC consolidated grant ST/P000681/1.

\end{document}